\documentclass[12pt]{article}

\begin{document}

\title{\bf Proper Matter Collineations of Plane Symmetric Spacetimes}

\author{M. Sharif \thanks{msharif@math.pu.edu.pk}\\
Department of Mathematics, University of the Punjab,\\
Quaid-e-Azam Campus, Lahore-54590, Pakistan\\
Tariq Ismaeel\\
Department of Mathematics, Government College University,\\
Lahore, Pakistan.}

\date{}

\maketitle

We investigate matter collineations of plane symmetric spacetimes
when the energy-momentum tensor is degenerate. There exists three
interesting cases where the group of matter collineations is
finite-dimensional. The matter collineations in these cases are
either {\it four}, {\it six} or {\it ten} in which {\it four} are
isometries and the rest are proper.\\

\date{}


Let $(M,g)$ be a spacetime, where $M$ is a smooth, connected,
Hausdorff four-dimensional manifold and $g$ is smooth Lorentzian
metric of signature (+ - - -) defined on $M$. The manifold $M$ and
the metric $g$ are assumed smooth ($C^{\infty}$). A smooth vector
field ${\bf \xi}$ is said to preserve a matter symmetry [1] on $M$
if, for each smooth local diffeomorphism $\phi_t$ associated with
${\bf \xi}$, the tensors $T$ and $\phi^*_tT$ are equal on the domain
$U$ of $\phi_t$, i.e., $T=\phi_t^*T$. Equivalently, a vector field
$\xi^a$ is said to generate a matter collineation if it satisfies
the following equation
\begin{equation}
\pounds_{\xi}T_{ab}=0,\quad  or\quad T_{ab,c} \xi^c + T_{ac}
\xi^c_{,b} + T_{cb} \xi^c_{,a}= 0,~~(a,b,c=0,1,2,3),
\end{equation}
where $\pounds$ is the Lie derivative operator, $\xi^a$ is the
symmetry or collineation vector. Every Killing vector (KV) is an MC
but the converse is not true, in general. A proper MC is an MC which
is not a KV, or a homothetic vector.

A large body of literature exists on classification of spacetimes
according to their isometries or Killing vectors and the groups
admitted by them [2-5]. These investigations of symmetries played an
important role in the classification of spacetimes, giving rise to
many interesting results with useful applications. As curvature and
Ricci tensors play a significant role in understanding the geometric
structure of metrics, the energy-momentum tensor enables us to
understand the physical structure of spacetimes. Some recent
investigations [6-19] show keen interest in the study of matter
collineations (MCs). In the papers [15-19], the study of MCs has
been taken for spherically symmetric, static plane symmetric and
cylindrically symmetric spacetimes and some interesting results have
been obtained. In this letter, we find the proper MCs of non-static
plane symmetric spacetimes when the energy-momentum tensor is
degenerate. It turns out that this admits an MC Lie algebra of 4, 6,
10 dimensions apart from the infinite dimensional algebras.

The most general form of the plane symmetric metric is given in the
form [4]
\begin{equation}
ds^2 =
e^{\nu(t,x)}dt^2-e^{\lambda(t,x)}dx^2-e^{\mu(t,x)}(dy^2+dz^2),
\end{equation}
where $\nu,~\lambda$ and $\mu$ are arbitrary functions of $t$ and
$x$. The surviving components of the energy-momentum tensor are
given as
\begin{eqnarray}
T_{00} &=&
\frac{1}{4}(\dot{\mu}^2+2\dot{\mu}\dot{\lambda})-\frac{1}{4}e^{v-\lambda}
(4\mu''+ 3 \mu'^2 - 2 \mu'\lambda'),\nonumber\\
T_{01}&=&-\frac{1}{2} (2 \dot{\mu}' + \dot{\mu}\mu' - \dot{\mu} v' -
\mu'\dot{ \lambda}), \nonumber \\
T_{11}&=&\frac{1}{4}({\mu'}^2+2\mu'\nu')-\frac{1}{4}e^{\lambda-\nu}
(4\ddot{\mu}+3\dot{\mu}^2-2\dot{\mu}\dot{\nu}),\nonumber\\
T_{22}& =&\frac{1}{4}e^{\mu-\nu}(2\mu''+{\mu'}^2
-\mu'\lambda'+\mu'\nu'-\lambda'\nu'+{\nu'}^2+2\nu'')\nonumber\\
&-&\frac{1}{4}e^{\mu-\nu}(2\ddot{\mu}+\dot{\mu}^2
-\dot{\mu}\dot{\nu}+\dot{\mu}\dot{\lambda}-\dot{\nu}\dot{\lambda
}+\dot{\lambda}^2+2\ddot{\lambda}),\nonumber \\
T_{33}& =& T_{22}.
\end{eqnarray}
The MC equations can be written as follows
\begin{eqnarray}
2T_{01}\xi^1_{,0}+T_{00,0} \xi^0+T_{00,1} \xi^1
+ 2 T_{00} \xi^0_{,0} &=& 0,\\
T_{01,0}\xi^0+T_{01}\xi^0_{0}+T_{00} \xi^0_{,1}+T_{01,1}\xi^1
+T_{11} \xi^1_{,0}+T_{01}\xi^1_{,1} &=& 0,\\
T_{00} \xi^0_{,2}+T_{01}\xi^1_{,2}+T_{22} \xi^2_{,0} &=& 0,\\
T_{00} \xi^0_{,3}+T_{01}\xi^1_{,3}+T_{22} \xi^3_{,0} &=&0,\\
T_{11,0} \xi^0+2T_{01}\xi^0_{1}+T_{11,1} \xi^1 + 2 T_{11} \xi^1_{,1} &= &0,\\
T_{01}\xi^0_{,2}+T_{11} \xi^1_{,2}+T_{22} \xi^2_{,1} &= &0,\\
T_{01}\xi^0_{,3}+T_{11} \xi^1_{,3}+T_{22}\xi^3_{,1}&=&0,\\
T_{22,0} \xi^0+T_{22,1} \xi^1 + 2 T_{22}\xi^2_{,2}&=&0,\\
T_{22}(\xi^2_{,3}+\xi^3_{,2}) &=&0,\\
T_{22,0} \xi^0+T_{22,1} \xi^1+ 2T_{22}\xi^3_{,3}& =& 0.
\end{eqnarray}
These are the first order non-linear partial differential equations
in four variables $\xi^a(x^b)$. We solve these equations for the
degenerate case, where $\det(T_{ab})=0$ and also we restrict
ourselves with the assumption $T_{01}=0$. For the sake of
simplicity, we use the notation $T_{aa}=T_a$. The following two
cases satisfy these assumptions:
\begin{description}
\item{(I)}$\quad SO(3)\otimes {\textbf{R}},$ where
${\textbf{R}}=\partial_t$ if
and only if\\
(a)$\quad \nu=\nu(x),~\lambda=\lambda(x),~\mu=2\ln x$ ~or\\
(b)$\quad
\nu=\nu(x),~\lambda=0,~\mu=2\ln a$,\\
$\quad$ where $a$ is an arbitrary constant.
\item{(II)}$\quad SO(3)\otimes {\textbf{R}},$ where
${\textbf{R}}=\partial_x$ if and only if\\
(a)$\quad \nu=\nu(t),~\lambda=\lambda(t),~\mu=2\ln t$ ~or\\
(b)$\quad \nu=0,~\lambda=\lambda(t),~\mu=2\ln a$.
\end{description}

\subsection*{CASE I}

This case has the following two possibilities:\\
(Ia)$\quad \nu=\nu(x),~\lambda=\lambda(x),~\mu=2\ln x$,\\
(Ib)$\quad \nu=\nu(x),~\lambda=0,~\mu=2\ln a.$\\
\textbf{Case (Ia)}: Here we solve MC Eqs.(4)-(13), when at least one
of $T_a=0$. This can be classified in the following three main
cases:
\begin{description}
\item[(1)] When only one of $T_a \neq 0$,
\item[(2)] When two of $T_a \neq 0$,
\item[(3)] When three of $T_a \neq 0$.
\end{description}
\textbf{Case (1)}: In this case, there could be further two
possibilities:
\begin{description}
\item[(1a)]$\quad T_0\neq 0,\quad T_{i}=0,\quad (i=1,2,3)$,
\item[(1b)]$\quad T_1\neq 0, \quad T_{j}=0,\quad (j=0,2,3)$.
\end{description}
When we solve MC equations of the case 1, we obtain infinite
dimensional MCs for all the possibilities.\\
\par\noindent
\textbf{Case (2)}: This case admits the following two subcases:
\begin{description}
\item[(2a)] $\quad T_{\ell} = 0, \quad T_k \neq0,\quad (\ell = 0,1)$,
\item[(2b)] $\quad T_{\ell} \neq0, \quad T_k = 0,\quad (k = 2,3)$.
\end{description}
After some algebra, we arrive at the conclusion that this case also
leads to infinite dimensional MCs in all the options.\\
\par\noindent
\textbf{Case (3)}: This is an interesting case of the degenerate
energy-momentum tensor which gives some finite dimensional MCs. Here
we take only one component of the energy-momentum tensor equals to
zero. The two possibilities are the following:
\begin{description}
\item [(3a)] $\quad T_0=0,\quad T_i\neq0,\quad(i=1,2,3).$
\item [(3b)] $\quad T_1=0,\quad T_j\neq0,\quad(j=0,2,3).$
\end{description}
\textbf{Case (3a)}: All the possibilities arising from this case
yield infinite dimensional MCs.\\
\par\noindent
\textbf{Case (3b)}: This case explores further four possibilities:
\begin{description}
\item [(i)] $\quad T_0=constant\neq 0,\quad T_2=constant\neq 0$,
\item [(ii)]$\quad T_{0,1}\neq 0,\quad T_2=constant\neq 0$,
\item [(iii)]$\quad T_0= constant\neq 0,\quad T_{2,1}\neq 0$,
\item [(iv)]$\quad T_{0,1}\neq 0,\quad T_{2,1}\neq 0$.
\end{description}
The cases 3b(i-iii) yield infinite dimensional MCs but the case
3b(iv) gives finite dimensional MCs. When we solve the MC equations
under the assumptions of this case, we further obtain the following
three options:
\begin{eqnarray*}
&(*)&\quad\frac{T_{0,1}T_2}{T_{2,1}T_0}=\varepsilon\neq 0,\quad
\left(\frac{T_0}{T_2}\right)'\neq0,\\
&(**)&\quad T_0=mT_2,\quad
(***)\quad\left(\frac{T_{0,1}T_2}{T_{2,1}T_0}\right)'\neq0,
\end{eqnarray*}
where $m$ is an arbitrary constant. In the first option, MCs turn
out to be
\begin{eqnarray}
\xi_{(1)}&=&\partial_t,\quad \xi_{(2)}=\partial_y,\quad
\xi_{(3)}=\partial_z,\quad \xi_{(4)}=z\partial_y-y\partial_z,\nonumber\\
\xi_{(5)}&=&t \partial_t-\frac{2T_0}{T_{0,1}}\partial_x,\quad
\xi_{(6)}=\iota (z\partial_y-y\partial_z).
\end{eqnarray}
This gives six independent MCs out of which two are proper MCs. The
Lie algebra has the following commutators:
\begin{eqnarray*}
[\xi_{(1)},\xi_{(5)}]=\xi_{(1)},\quad
[\xi_{(2)},\xi_{(4)}]=-\xi_{(3)},\quad
[\xi_{(2)},\xi_{(6)}]=-\iota\xi_{(3)},
\end{eqnarray*}
\begin{eqnarray*} [\xi_{(3)},\xi_{(4)}]=\xi_{(2)},\quad
[\xi_{(3)},\xi_{(6)}]=\iota\xi_{(2)},\quad[\xi_{(i)},\xi_{(j)}]=0,~\textrm{(otherwise).}
\end{eqnarray*}
For the second possibility, we obtain six proper MCs given by
\begin{eqnarray}
\xi_{(5)}&=&\frac{1}{m}\left[ ty\partial_t-\frac{2T_2}{T_{2,1}}y
\partial_x+\frac{1}{2}(y^2-z^2-m
t^2)\partial_y+yz \partial_z\right],\nonumber\\
\xi_{(6)}&=&tz\partial_t-\frac{2T_2}{T_{2,1}}z\partial_x+yz
\partial_y-\frac{1}{2}(y^2-z^2-m
t^2)\partial_z,\nonumber\\
\xi_{(7)}&=&t\partial_t-\frac{2T_2}{T_{2,1}}
\partial_x+y\partial_y+z \partial_z,\nonumber\\
\xi_{(8)}&=&\frac{1}{2m}(y^2+z^2-m t^2)\partial_t
+\frac{2T_2}{T_{2,1}}t \partial_x-ty \partial_y-tz \partial_z,\nonumber\\
\xi_{(9)}&=&\frac{1}{m}y \partial_t-t\partial_y,\quad
\xi_{(10)}=\frac{1}{m}z \partial_t-t\partial_z.
\end{eqnarray}
The Lie algebra is given by
\begin{eqnarray*}
[\xi_{(1)},\xi_{(5)}]=\xi_{(9)},\quad
[\xi_{(1)},\xi_{(6)}]=m
\xi_{(10)},\quad [\xi_{(1)},\xi_{(7)}]=\xi_{(1)},
\end{eqnarray*}
\begin{eqnarray*}
[\xi_{(1)},\xi_{(8)}]=-\xi_{(7)},\quad
[\xi_{(1)},\xi_{(9)}]=-\xi_{(2)},\quad
[\xi_{(1)},\xi_{(10)}]=-\xi_{(3)},
\end{eqnarray*}
\begin{eqnarray*}
[\xi_{(2)},\xi_{(4)}]=-\xi_{(3)},\quad
[\xi_{(2)},\xi_{(5)}]=\frac{1}{m}\xi_{(7)},\quad
[\xi_{(2)},\xi_{(6)}]=\xi_{(4)},
\end{eqnarray*}
\begin{eqnarray*}
[\xi_{(2)},\xi_{(7)}]=\xi_{(2)},\quad
[\xi_{(2)},\xi_{(8)}]=\xi_{(9)},\quad
[\xi_{(2)},\xi_{(9)}]=\frac{1}{m}\xi_{(1)},
\end{eqnarray*}
\begin{eqnarray*}
[\xi_{(3)},\xi_{(4)}]=\xi_{(2)},\quad
[\xi_{(3)},\xi_{(5)}]=-\frac{1}{m}\xi_{(4)},\quad
[\xi_{(3)},\xi_{(6)}]=\xi_{(7)},
\end{eqnarray*}
\begin{eqnarray*}
[\xi_{(3)},\xi_{(7)}]=\xi_{(3)}, \quad
[\xi_{(3)},\xi_{(8)}]=\xi_{(10)},\quad
[\xi_{(3)},\xi_{(10)}]=\frac{1}{m}\xi_{(1)},
\end{eqnarray*}
\begin{eqnarray*}
[\xi_{(4)},\xi_{(5)}]=\frac{1}{m}\xi_{(6)}, \quad
[\xi_{(4)},\xi_{(6)}]=-m\xi_{(5)},\quad
[\xi_{(4)},\xi_{(7)}]=\xi_{(4)},
\end{eqnarray*}
\begin{eqnarray*}
[\xi_{(4)},\xi_{(10)}]=-\xi_{(9)},\quad
[\xi_{(5)},\xi_{(7)}]=\xi_{(5)},\quad
[\xi_{(5)},\xi_{(9)}]=-\frac{1}{m}\xi_{(8)},
\end{eqnarray*}
\begin{eqnarray*}
[\xi_{(6)},\xi_{(7)}]=\xi_{(6)},\quad
[\xi_{(6)},\xi_{(10)}]=-\xi_{(8)},\quad
[\xi_{(7)},\xi_{(8)}]=\xi_{(8)},
\end{eqnarray*}
\begin{eqnarray*}
[\xi_{(8)},\xi_{(9)}]=\xi_{(5)},\quad
[\xi_{(8)},\xi_{(10)}]=\frac{1}{m}\xi_{(6)},\quad
[\xi_{(9)},\xi_{(10)}]=\frac{1}{m}\xi_{(4)},
\end{eqnarray*}
\begin{eqnarray*}
[\xi_{(i)},\xi_{(j)}]&=&0,\quad \textrm{(otherwise).}
\end{eqnarray*}
The last option yields four independent MCs which are the usual KVs
of the static plane symmetry.

\subsection*{CASE II}

This case has two subcases\\
(a) $\quad \nu=\nu(t),~\lambda=\lambda(t),~\mu=2\ln t$,\\
(b)$\quad \nu=0,~\lambda=\lambda(t),~\mu=2\ln a$.\\
\par\noindent
\textbf{Case(IIa)}: The other possibilities can be classified in the
following three main cases:
\begin{description}
\item[(1)] When only one of $T_a \neq 0$,
\item[(2)] When two of $T_a \neq 0$,
\item[(3)] When three of $T_a \neq 0$.
\end{description}
The cases (1) and (2) lead to infinite dimensional MCs in
all the possibilities.\\
\par\noindent
\textbf{Case (3)}: This case of the degenerate energy-momentum
tensor yields some finite dimensional MCs. The following two
possibilities arise:
\begin{description}
\item [(3a)] $\quad T_0=0,\quad T_i\neq0,\quad\ (i=1,2,3).$
\item [(3b)] $\quad T_1=0,\quad T_j\neq0,\quad\ (j=1,2,3).$
\end{description}
\textbf{Case (3a)}: This case further leads to the following four
possibilities:
\begin{description}
\item [(i)]$\quad T_1=constant\neq 0,\quad T_2=constant\neq 0,$
\item [(ii)]$\quad T_{1,0}\neq 0,\quad T_2=constant\neq 0,$
\item [(iii)]$\quad T_1=constant\neq 0,\quad T_{2,0}\neq 0,$
\item [(iv)]$\quad T_{1,0}\neq 0,\quad T_{2,0}\neq 0$.
\end{description}
The cases 3a(i) and 3a(ii) yield the infinite dimensional MCs.\\
\par\noindent
\textbf{Case (3aiii)}: When we solve MC equations simultaneously for
this case, we obtain the following finite dimensional MCs
\begin{eqnarray}
\xi_{(1)}&=&\partial_x,\quad \xi_{(2)}=\partial_y,\quad
\xi_{(3)}=\partial_z,\quad
\xi_{(4)}=z\partial_y-y\partial_z,\nonumber\\
\xi_{(5)}&=&y\partial_x-m x\partial_y,\quad \xi_{(6)}=z\partial_x-m
x\partial_z.
\end{eqnarray}
We obtain six independent MCs. The corresponding Lie algebra has the
following commutators:
\begin{eqnarray*}
[\xi_{(1)},\xi_{(6)}]=-m\xi_{(3)},\quad
[\xi_{(1)},\xi_{(5)}]=-m\xi_{(2)},\quad
[\xi_{(6)},\xi_{(5)}]=-m\xi_{(4)},
\end{eqnarray*}
\begin{eqnarray*}
[\xi_{(6)},\xi_{(4)}]=-\xi_{(5)},\quad
[\xi_{(6)},\xi_{(3)}]=-\xi_{(1)},\quad[\xi_{(5)},\xi_{(4)}]=-\xi_{(6)},
\end{eqnarray*}
\begin{eqnarray*}
[\xi_{(5)},\xi_{(2)}]=-\xi_{(1)},\quad
[\xi_{(4)},\xi_{(2)}]=\xi_{(3)},\quad
[\xi_{(4)},\xi_{(3)}]=-\xi_{(2)},
\end{eqnarray*}
\begin{eqnarray*}
[\xi_{(i)},\xi_{(j)}]&=&0,\quad \textrm{(otherwise).}
\end{eqnarray*}
\textbf{Case (3aiv)}: This case turns out exactly the same as
(3aiii).\\
\textbf{Case (3b)}: This case leads to infinite dimensional MCs.\\
\textbf{Case (IIb)}: It is similar to the case II(2a).


In a recent paper [19], some interesting results have been obtained
when we classify static plane symmetric spacetimes according to
their energy-momentum tensor. This idea has been used to find the
proper MCs of non-static plane symmetric spacetimes for the
degenerate case only. It is worth mentioning that we have found four
such cases having either four, six or ten independent MCs. The
results are summarized in the form of table 1.

\vspace{0.4cm}

{\bf {\small Table 1}. }{\small MCs for the Degenerate Case (only
finite cases)}

\vspace{0.1cm}

\begin{center}
\begin{tabular}{|l|l|l|}
\hline {\bf Cases} & {\bf MCs} & {\bf Constraints}
\\ \hline I3biv(*) & $6$ & $T_1=0,~T_j\neq 0(j=0,2,3),~T'_2\neq 0,~
\frac{T'_0T_2}{T_0T'_2}\neq 0,~(\frac{T_0}{T_2})'\neq 0$
\\ \hline I3biv(**) & $10$ &
$T_1=0,~T_j\neq 0,~T'_2\neq 0,~T_0=m T_2,~T'_0\neq 0$
\\ \hline I3biv(***) & $4$ &
$T_1=0,~T_j\neq 0,~T'_2\neq 0,~\frac{T'_0T_2}{T_0T'_2}\neq 0$
\\ \hline II3a(iii) & $6$ &
$T_0=0,~T_{1,0}\neq 0,~T_{2,0}\neq 0,~T_1=mT_2$
\\ \hline
\end{tabular}
\end{center}
This shows that each case has different constraints on the
energy-momentum tensor. When the rank of $T_a$ is 3, i.e. $T_1=0$,
we obtain the following metric
\begin{equation}
ds^2=e^\nu dt^2-dx^2-e^{-2\nu}(dy^2+dz^2),
\end{equation}
where $\nu$ is an arbitrary function of $x$ only. It can be easily
verified that this class of metrics represent perfect fluid dust
solutions. The energy-density for the above metrics is given as
\begin{equation}
\rho=(2\nu''-3\nu'^2)e^{\frac{\nu}{2}},
\end{equation}
We can conclude that when the rank of $T_{ab}$ is 1 or 2, all the
possibilities yield infinite dimensional MCs. If the rank of
$T_{ab}$ is 3, this leads to some possibilities of finite
dimensional MCs. It would be interesting to classify plane symmetric
spacetimes according to their MCs for the non-degenerate case and
then removing the assumption $T_{01}=0$. All these problems would be
investigated as a separate issue.

\newpage

{\bf \large References}

\begin{description}

\item{[1]} Hall, G.S.: Gen. Rel and Grav. {\bf 30}(1998)1099;
\emph{Symmetries and Curvature Structure in General Relativity}
(World Scientific, 2004).

\item{[2]} Katzin, G.H., Levine J. and Davis, W.R.: J. Math. Phys.
{\bf 10}(1969)617.

\item{[3]} Petrov, A.Z.: {\it Einstein Spaces} (Pergamon, Oxford
University Press, 1969).

\item{[4]} Stephani, H., Kramer, D., MacCallum, M.A.H.,
Hoenselaers, C. and Hearlt, E.: {\it Exact Solutions of Einstein's
Field Equations} (Cambridge University Press, 2003).

\item{[5]} Rcheulishrili, G.: J. Math. Phys. {\bf 33}(1992)1103.

\item{[6]} Hall, G.S., Roy, I. and Vaz, L.R.: Gen. Rel and Grav.
{\bf 28}(1996)299.

\item{[7]} Camc{\i}, U. and Barnes, A.: Class. Quant. Grav. {\bf
19}(2002)393.

\item{[8]} Carot, J. and da Costa, J.: {\it Procs. of the 6th
Canadian Conf. on General Relativity and Relativistic Astrophysics},
Fields Inst. Commun. 15, Amer. Math. Soc. WC Providence,
RI(1997)179.

\item{[9]} Carot, J., da Costa, J. and Vaz, E.G.L.R.: J. Math. Phys. {\bf
35}(1994)4832.

\item{[10]} Tsamparlis, M., and Apostolopoulos, P.S.: J. Math.
Phys. {\bf 41}(2000)7543.

\item{[11]} Sharif, M.: Nuovo Cimento {\bf B116}(2001)673.

\item{[12]} Astrophys. Space Sci. {\bf 278}(2001)447.

\item{[13]} Camc{\i}, U. and Sharif, M.: Gen Rel. and Grav. {\bf
35}(2003)97.

\item{[14]} Camc{\i}, U. and Sharif, M.: Class. Quant. Grav. {\bf 20}(2003)2169.

\item{[15]} Sharif, M. and Aziz, Sehar: Gen Rel. and Grav. {\bf
35}(2003)1091.

\item{[16]} Sharif, M.: J. Math. Phys. {\bf 44}(2003)5141.

\item{[17]} Sharif, M.: J. Math. Phys. {\bf
45}(2004)1518.

\item{[18]} Sharif, M.: J. Math. Phys. {\bf 45}(2004)1532.

\item{[19]} Sharif, M.: J. Math. Phys. {\bf 45}(2004)1518.

\end{description}

\end{document}